\documentclass[ twocolumn, showkeys,prb,aps,superscriptaddress,preprintnumbers,amsmath,amssymb]{revtex4-1}

\usepackage{graphicx}% Include figure files
\usepackage{dcolumn}% Align table columns on decimal point
\usepackage{bm}% bold math
\usepackage{multirow}
\usepackage[breaklinks=true,colorlinks,citecolor=blue,linkcolor=blue,urlcolor=blue]{hyperref}
\usepackage{mathtools}
\usepackage{braket}
\usepackage{color}
\usepackage{url}
\usepackage{epstopdf}
\usepackage{amssymb}
\usepackage{amsmath}
\def\CAB{${\rm CaAgBi}$}
\def\A100{[10\={1}0] }
\def\B120{[1\={2}10]}
\begin{document}

\title{Weak antilocalization in a noncentrosymmetric ${\rm CaAgBi}$ single crystal}%
\author{Souvik Sasmal}
\affiliation{Department of Condensed Matter Physics and Materials Science, Tata Institute of Fundamental Research, Homi Bhabha Road, Colaba, Mumbai 400 005, India.}

\author{Rajib Mondal}
\affiliation{Department of Condensed Matter Physics and Materials Science, Tata Institute of Fundamental Research, Homi Bhabha Road, Colaba, Mumbai 400 005, India.}

\author{Ruta Kulkarni}
\affiliation{Department of Condensed Matter Physics and Materials Science, Tata Institute of Fundamental Research, Homi Bhabha Road, Colaba, Mumbai 400 005, India.}

\author{Bahadur Singh}
\affiliation{Department of Physics, Northeastern University, Boston, Massachusetts 02115, USA}
\affiliation{SZU-NUS Collaborative Center and International Collaborative Laboratory of 2D Materials for Optoelectronic Science and Technology, College of Optoelectronic Engineering, Shenzhen University, Shenzhen 518060, China}

\author{A. Thamizhavel}
\affiliation{Department of Condensed Matter Physics and Materials Science, Tata Institute of Fundamental Research, Homi Bhabha Road, Colaba, Mumbai 400 005, India.}

\date{\today}

\begin{abstract}

We report on the single crystal growth and transport properties of a topological semimetal~ \CAB ~which crystallizes in the hexagonal $ABC-$type structure with the non-centrosymmetric space group $\mathit{P6_3mc}$ (No. 186). The transverse magnetoresistance measurements with current in the basal plane of the hexagonal crystal structure reveal a value of about 30~\% for $I~\parallel$~[10\={1}0] direction and about 50~\% for $I~\parallel$~[1\={2}10] direction at 10~K in an applied magnetic field of 14~T. The magnetoresistance shows a cusp-like behavior in the low magnetic field region, suggesting the presence of weak antilocalization effect for temperatures less than 100~K. The Hall measurements reveal that predominant charge carriers are $p$-type exhibiting a linear behavior for fields up to 14 T and can be explained based on the single band model. The magnetoconductance of \CAB~ is analysed based on the modified Hikami-Larkin-Nagaoka (HLN) model. Our first-principles calculations within a density-functional theory framework reveal that ~\CAB~ supports a topological Dirac semimetal state with Dirac points located on the rotational axis slightly above the Fermi level and are protected by $C_{6v}$ point-group symmetry. The Fermi surface consists of both the electron and hole pockets. However, the size of hole pockets is much larger than electron pockets suggesting the dominant $p$-type carriers in accord with our experimental results. 

%The strong spin-orbit coupling in \CAB~ results in a cusp-like behavior in the low field region of the magnetoresistance, suggesting weak-antilocalization for temperatures less than 100~K. 

\end{abstract}

%\pacs{81.10.-h, 71.70.Ch, 75.10.Dg, 75.50.Gg, 71.70.Gm, 75.30.Sg} 

\maketitle

\section{Introduction}

Symmetry-protected non-trivial states in topological insulators and semimetals have opened a flood gate of research activities in the condensed matter physics owing to their interesting physical properties that are useful not only for fundamental research but also for next-generation electronic/spintronic device applications~\cite{RevModPhys.88.021004,Burkov2016, Kong2011}. The topological semimetals are mainly classified as Weyl semimetal, Dirac semimetal, nodal-line semimetal (NLSM), and unconventional semimetal, like a triply degenerate nodal semimetal~\cite{PhysRevB.83.205101,Weng_2016, PhysRevLett.115.036806, doi:10.1146/annurev-matsci-070218-010049, PhysRevX.5.011029, Bradlynaaf5037, PhysRevB.95.195165, PhysRevB.93.241202, PhysRevLett.121.106404,PhysRevMaterials.1.044201,PhysRevLett.108.140405,Yang2014, PhysRevMaterials.3.071201, PhysRevB.84.235126}. In a Weyl, Dirac, or unconventional semimetal, the valence and conduction bands form discrete two-, four-, and higher-fold degenerate crossing points in the Brillouin zone. On the contrary, the valence and conduction bands crossings in NLSM form a line or loop in the momentum space near Fermi level. The band crossings in a Dirac semimetal can be regarded as a superposition of two Weyl fermions of opposite chiral charge in presence of time-reversal ($\mathcal{T})$ and inversion ($\mathcal{I}$) symmetry. These band crossings are protected against band hybridization by additional crystalline symmetries. Breaking of either $\mathcal{T}$, $\mathcal{I}$, or both the symmetries leads to a Weyl semimetal state where two Weyl fermions of opposite chiral charge are separated in the momentum space. The topological semimetals harbor exotic phenomena such as extremely large magnetoresistance~\cite{PhysRevB.96.075159, PhysRevB.97.205130}, quantum oscillation~\cite{Pavlosiuk2016}, quantum coherence effect~\cite{PhysRevLett.106.166805, PhysRevLett.108.036805}, non-linear Hall effect~\cite{Kang2019}, among others \cite{PhysRevX.5.031023,Chang2018}.

The NLSMs state has been theoretically predicted in many class of materials including elemental alkali metals where the band crossings form closed Dirac nodal lines around the $\Gamma$ point~\cite{PhysRevLett.117.096401}. However, the experimental evidence of NLSM state has been reported in Mg$_3$Bi$_2$~\cite{doi:10.1002/advs.201800897}, PbTaSe$_2$~\cite{Bian2016}, ZrSiS~\cite{doi:10.1021/acs.nanolett.7b02307, Schoop2016}, CaAgAs~\cite{Nayak_2018} materials. Studies on polycrystalline and single crystals of ${\rm CaAgAs}$ have revealed a doughnut-like Fermi surface suggesting an ideal candidate material for studying the NLSMs. More recently, it has been predicted that the hexagonal $ABC-$type ${\rm SrHgPb}$ class of materials with polar space group $P6_3mc$ ($\# 186$) have both the Dirac and Weyl points in their electronic structure~\cite{PhysRevLett.121.106404}. It is, therefore, worthwhile to synthesize and investigate the transport properties of hexagonal $ABC-$type topological materials which can facilitate the discovery of Dirac-Weyl semimetals in polar materials~\cite{PhysRevLett.121.106404,PhysRevMaterials.1.044201}.

The spin-momentum locking caused by strong spin-orbit coupling (SOC) in Dirac-cone states of topological insulators gives rise to a $\pi$ Berry phase which results in a reduction in backscattering and causes quantum interference~\cite{Liu_2015, 10.1117/12.2063426, Xu2014}. In a weak disorder driven quantum diffusive regime if phase coherence length ($L_{\rm \phi}$) is much greater than the mean free path $(l)$, the electrons maintain phase coherence even after being elastically scattered many times so that the time-reversed interference path causes weak localization (WL) and weak antilocalization (WAL)~\cite{10.1117/12.2063426}. In the case of WL, the constructive interference suppresses the conductivity while in WAL the destructive interference enhances the conductivity at low temperature~\cite{10.1117/12.2063426, datta_1995}. Application of a small magnetic field destroys the interference and a cusp-like positive and negative conductivity appears as a result of WL and WAL. The WL and WAL effects can be observed dominantly in the lower-dimensional systems because of the higher probability of scattering events occurring between two time-reversed paths resulting in the interference effect. The WAL effects have also been reported in single crystals of different materials~\cite{PhysRevB.95.195113,PhysRevB.99.241102}. In this work, we report on the basal plane anisotropy in the transport properties of topological semimetal \CAB~ single crystals grown by the self-flux method and show that it exhibits WAL in the magnetoconductivity.

\section{Methods}\label{methods}

Single crystals of \CAB~were grown by the self-flux method. High purity starting elements of Ca (99.8\%), Ag (99.99\%) and Bi (99.999\%) from M/s. Alfa Aesar were taken in the molar ratio $1:2:20$ in a high quality recrystallized alumina crucible. Here the excess Ag was necessary to obtain single crystals of CaAgBi. The crucible was placed in a quartz ampoule and evacuated to a vacuum of $10^{-6}$~Torr and finally sealed under a partial pressure of Ar gas. The sealed ampoule was placed in a box-type resistive heating furnace and heated at the rate of 50~$^{\circ}$C/h to 1050~$^{\circ}$C and held at this temperature for 24~hrs homogenization. The furnace was then cooled down to 420~$^{\circ}$C at the rate of 2~$^{\circ}$C/h at which point the excess Bi flux was centrifuged. Several flat shiny crystals with a typical dimension of $3 \times 2 \times 2$~mm$^{3}$ were obtained. The grown crystals were stable in air for an extended period of time. Powder x-ray and Laue diffraction experiments were performed to find the phase purity and to determine the crystallographic orientation. Clean single crystals were carefully selected and cut them into thin bar shaped sample to make sure that the residual bismuth is completely removed from the crystal for our electrical resistivity measurements. The rectangular bar-shaped single crystals with a length corresponding to \A100 and \B120 directions were cut for the electrical transport property measurements. The electrical resistivity and Hall effect measurements were measured using a Quantum Design Physical Property Measurement System (PPMS) equipped with a 14~T magnet. 

Electronic structure calculations were performed within the framework of density functional theory (DFT)\cite{PhysRev.136.B864}, using the Vienna {\it ab-initio} simulation package (VASP)\cite{vasp}. We used the projector augmented wave (PAW) method to treat the interaction between the ion cores and valence electrons and generalized gradient approximation (GGA) to consider exchange-correlation effects\cite{paw,gga}. The SOC was included self-consistently to include relativistic effects. A plane wave-cut off energy of 420 eV was employed and a $9\times9\times7$ $\Gamma$-centered $k$ mesh is used for bulk computations. The topological properties were calculated by employing a tight-binding model Hamiltonian using WANNIERTOOLS package\cite{PhysRevB.56.12847,wanniertools}. Ag $s$ and Bi $p$ states were considered to obtain the tight-binding model via VASP2WANNIER90 interface. The experimental structural parameters were employed to calculate electronic properties.  

%Buckling of Hg-Pb layer in the ${\rm SrHgPb}$ crystal structure can break inversion symmetry and can establish co-existence of both distinct type of topological phases. 

\section{Results and Discussions}

\subsection{X-ray diffraction}

\begin{figure}[!]
\includegraphics[width=0.5\textwidth]{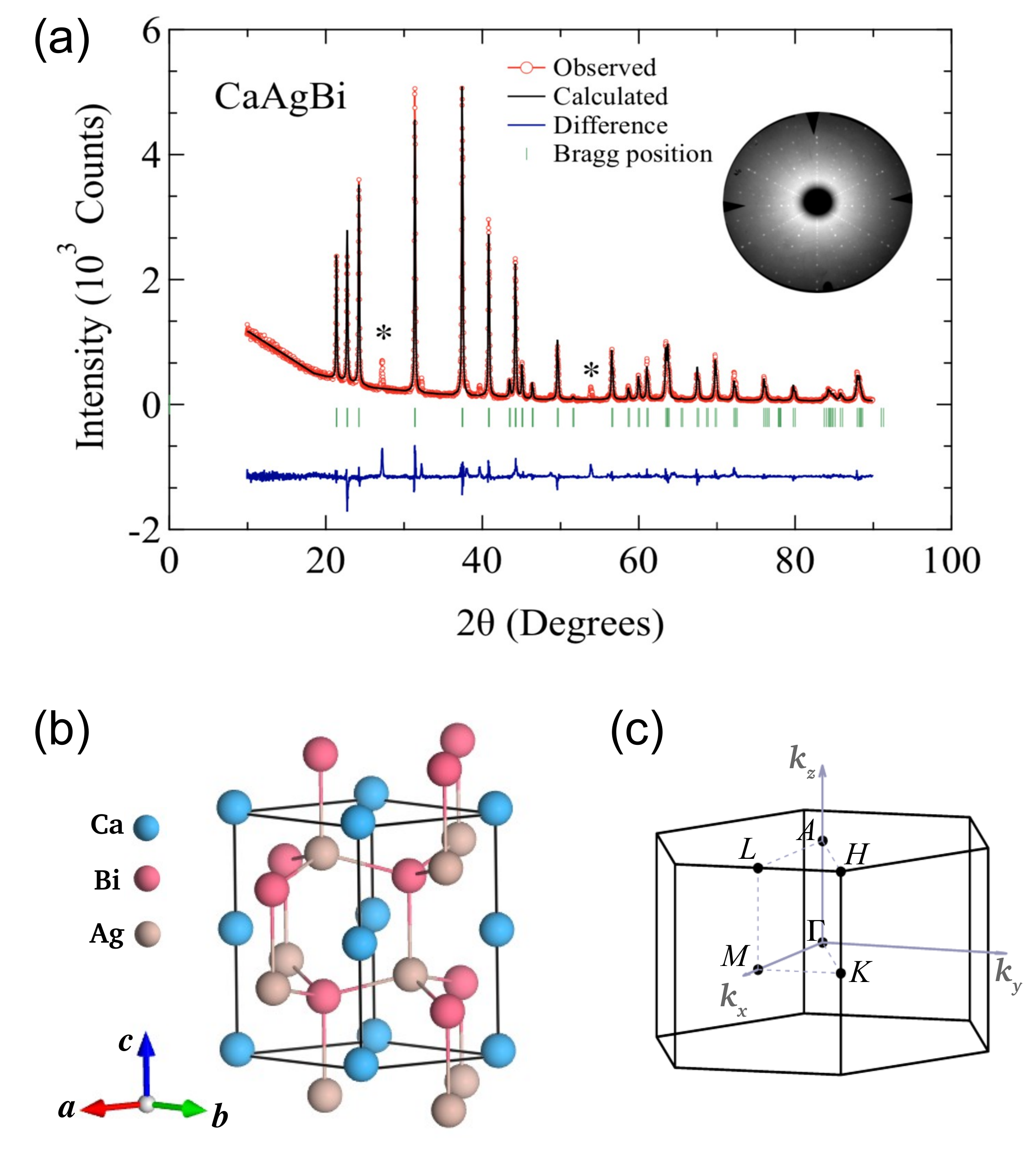}
\caption{(Color online) Powder x-ray diffraction pattern of crushed single crystals of \CAB. Inset shows the Laue diffraction pattern corresponding to the (0001) plane. (b) Hexagonal crystal structure of \CAB. The buckling of Ag and Bi is also shown. (c) The corresponding bulk Brillouin zone. Various high-symmetry points are marked. }
\label{Fig1}
\end{figure}

To check the phase purity of the grown \CAB~single crystals, we have performed the powder x-ray diffraction (XRD) analysis by grinding a few pieces of the single crystals. The XRD measurement was performed using Cu-$K_{\rm \alpha}$ source with wavelength $\lambda = 1.5406$~\AA. Figure ~\ref{Fig1}(a) shows the powder XRD pattern measured at 300~K. It is obvious from the XRD pattern that as grown crystals consist of a single phase and the impurity peaks observed at around 27$^{\circ}$ and 55$^{\circ}$ 2$\theta$ are due to the trace amounts of bismuth flux. From the Rietveld analysis, using the FULLPROF software~\cite{Rietveld:a07067, Rodriguez1990}, of the XRD pattern we confirmed that this compound crystallizes in the hexagonal crystal structure with non-centrosymmetric space group ${P6_3mc}$ (\#186) [see Fig. \ref{Fig1}(b) and details below]. The estimated lattice constants $a = 4.805$~\AA~ and $c = 7.828$~\AA~ match well with the previously published data~\cite{SUN200771}.  The Laue diffraction in the back-reflection geometry has been performed on the flat plane of the single crystal which corresponds to (0001) and is shown in the inset of Fig.~\ref{Fig1}(a).  Since the grown crystals were very thin we could not make the current contacts along [0001] direction. A 50~$\mu$m gold wire was used for the four-probe electrical resistivity and Hall measurements. EPOTEK H20E silver epoxy was used for making the contacts.

%*********************FIGURE 1********************************
%****************************************************************
Figure \ref{Fig1}(b) illustrates the crystal structure of \CAB~ in non-centrosymmetric hexagonal space group ${P6_3mc}$. It forms a stuffed wurtzite lattice where the buckled Ag and Bi atoms with Wyckoff positions $2b$ form the wurtzite lattice whereas the Ca atoms with Wyckoff position $2a$ occupy the interstitial sites in the AgBi lattice. The buckling in the Ag and Bi atomic layers preserves the $M_{110}$ mirror plane $M_x$ (we choose [110] direction as $x$) but breaks inversion symmetry. The other important symmetries include three-fold rotation $C_{3z}$ and two-fold screw rotation $S_{2z}=\{C_{2z}|00\frac{1}{2}\}$ symmetries. Additionally, the crystal respects the time-reversal symmetry $\mathcal{T}$. The bulk Brillouin zone is shown in Fig. \ref{Fig1}(c) where high-symmetry points are marked. 

\subsection{Electrical Resistivity}

%*********************FIGURE 2********************************
\begin{figure}[!]
\includegraphics[width=0.5\textwidth]{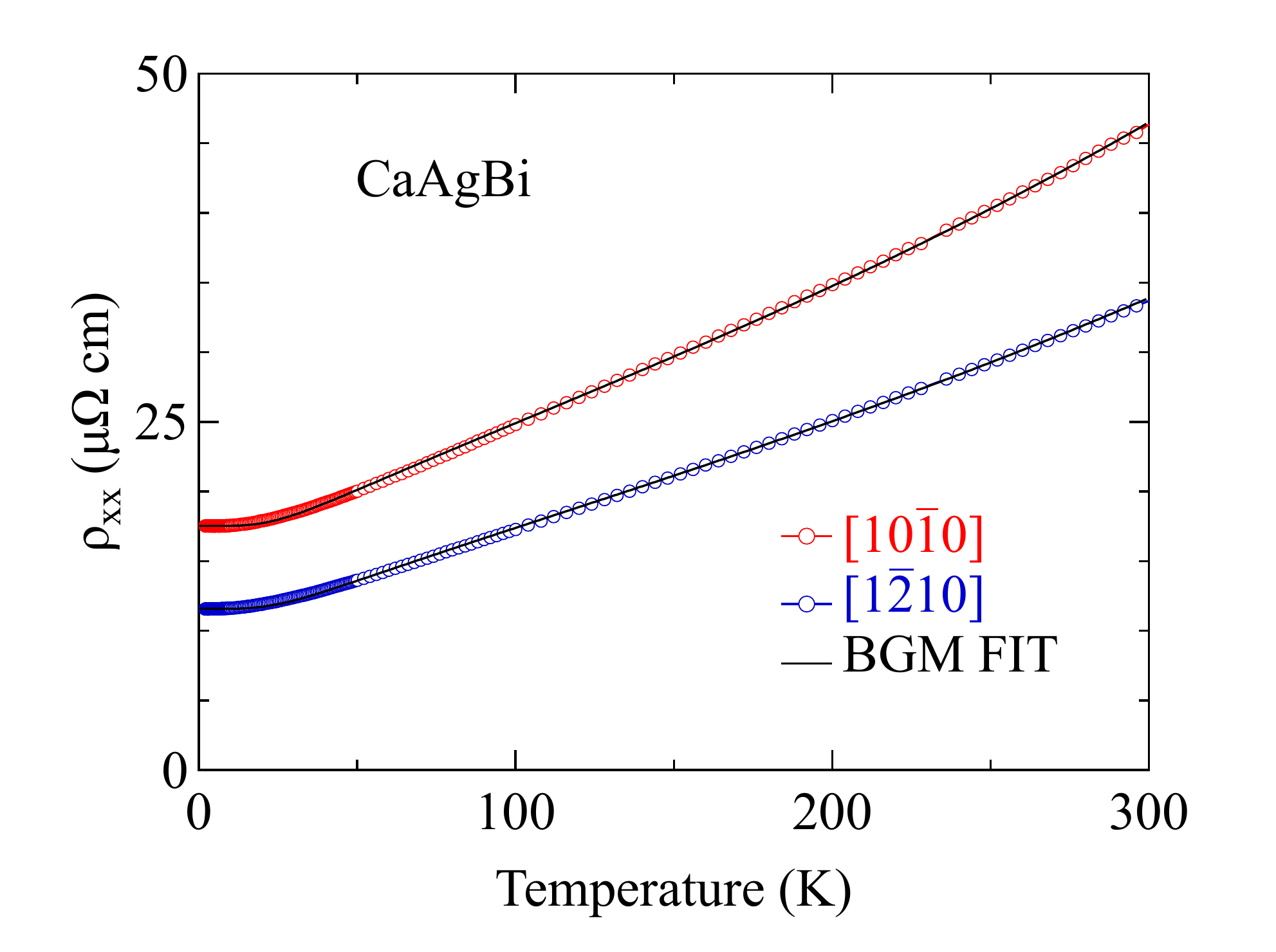}
\caption{(Color online) (a) Temperature dependence of electrical resistivity of \CAB~measured in the $ab$-plane for current parallel to \A100 and \B120 directions, respectively. The solid lines are the fit to the Bloch-Gruneissen-Mott (BGM) expression (see text for details). }
\label{Fig2}
\end{figure}
%****************************************************************
The zero-field electrical resistivity of CaAgBi measured along basal plane directions \A100 and \B120 in the temperature range 2 to 300~K are shown in Fig.~\ref{Fig2}. The overall behavior of the resistivity is similar to that of the previously reported data on a polycrystalline sample~\cite{MERLO1995280}. There is a very subtle anisotropy in the electrical resistivity. At 300~K the electrical resistivity is 46.2~$\mu \Omega$ cm and 33.6~$\mu \Omega$ cm along \A100 and \B120~ directions, respectively. The resistivity decreases with decreasing temperature owing to the reduction in the electron-phonon scatterings and levels off at around 20~K. It becomes temperature independent as the temperature is lowered down, as expected for normal metals without any strong electron correlations. The resistivity follows the Bloch-Gr\"{u}neisen-Mott (BGM) expression which is given by:

%***********************Equation 1 ********************
\begin{equation}
\label{Eqn1}
\rho = \rho_0 + 4 R T \left( \frac{T}{\theta_{\rm R}} \right)^4 J_5\left(\frac{\theta_R}{T}\right) + KT^3,
\end{equation}

where $J_5(\theta_R/T)$ is the Gr\"{u}neisen function. Here the first term $\rho_0$ is the residual resistivity, the second term corresponds to electron-phonon scattering, and the final $T^3$ term represents the contribution due to Mott's $s-d$ interband electron scattering. A least-square fit to the electrical resistivity of CaAgBi resulted in the following values: $\rho_0$ = 17.54~$\mu \Omega$cm, $R = 0.081$~$\mu \Omega$cm/K and $K = 1.81~\times~10^{-7}$~$\mu \Omega$cm/K$^3$ for J~//~\A100 and $\rho_0$ = 11.58~$\mu \Omega$cm, $R = 0.064$~$\mu \Omega$cm/K and $K = 1.18~\times~10^{-7}$~$\mu \Omega$cm/K$^3$ for J~//~\B120.  One of the reasons for the relatively high value of the electrical resistivity at low temperature might be presence of site-disorder between Bi and Ag atoms as both of them occupy the same Wyckoff positions $2b$.

 \subsection{Magnetoresistance and Hall effect}

%*********************FIGURE 3********************************
\begin{figure}
\includegraphics[width=0.5\textwidth]{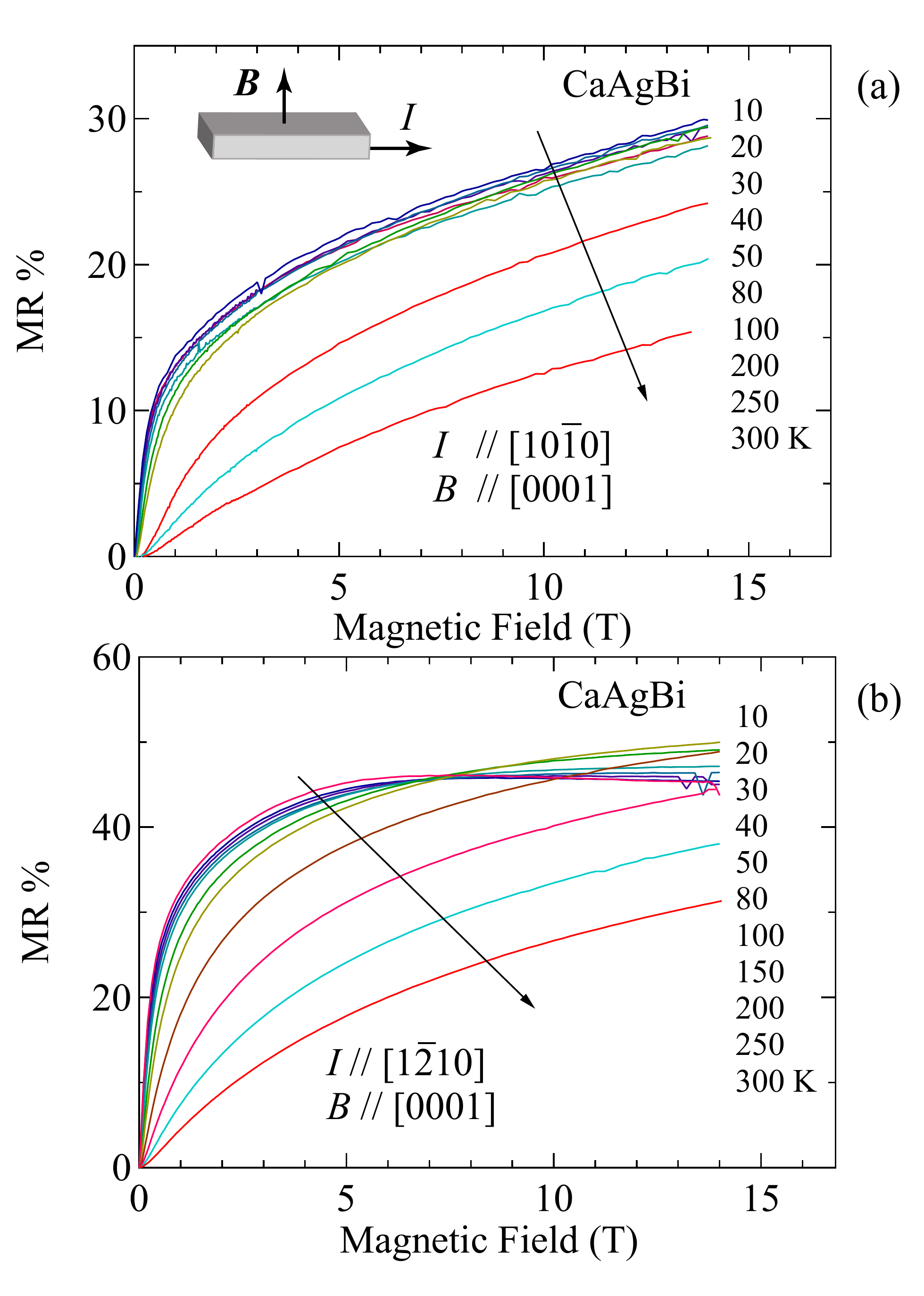}
\caption{(Color online) (a) Magnetoresistance (MR) of \CAB~ as a function of magnetic field at selected temperatures for $I~\parallel~$\A100 and $B~\parallel$~[0001]. MR is defined as $[(\rho (H) - \rho (0))/\rho (0)]~\times~100\%$. (b)  MR for $I~\parallel~$\B120 and $B~\parallel~[0001]$ direction. }
\label{Fig3}
\end{figure}
%****************************************************************

The magnetic field dependence of magnetoresistance for current along the two principal directions in the basal plane and field parallel to the [0001] direction at selected temperatures are shown in Figs.~\ref{Fig3}(a) and \ref{Fig3}(b). For current parallel to \A100 and temperatures less than 100~K, the MR at low fields increases very rapidly with increasing field. For magnetic fields greater than $5$~T, it increases linearly without showing any sign of saturation. The MR reaches almost 30~\% at $14$~T as shown in Fig.~\ref{Fig3}(a). On the other hand, for current parallel to \B120 the MR increases very rapidly as the field is increased remains almost flat for temperature less than $50$~K for fields higher than $5$~T, and for temperatures greater than $50$~K, the MR increases linearly for fields greater than $5$~T. The MR reaches 50~\% at $14$~T for current parallel to \B120 as shown in Fig.~\ref{Fig3}(b).  The low value of the magnetoresistance compared to other Dirac or Weyl semimetals like  NbP~\cite{Shekhar2015}, Cd$_3$As$_2$~\cite{Liang2014}, or MoSi$_2$~\cite{PhysRevB.97.205130} is mainly attributed to the low carrier mobility in CaAgBi, as observed from Hall data, to be discussed later. A rather sharp cusp-like behavior is observed in the low-field region of the MR for temperature less than $100$~K in both the principal directions of the $ab$-plane. This resembles the weak antilocalization (WAL) effects which have been typically observed in 2D-systems~\cite{PhysRevLett.103.246601, Bao2012} and as well in some of the Bi-based topological half-Heusler alloy systems~\cite{PhysRevB.94.035130, doi:10.1063/1.4936179, Xu2014}. The WAL effects arise when the probability of electron localization gets reduced due to the destructive interference of the electron wave functions in the two time-reversed electron paths~\cite{datta_1995, BERGMANN19841} and also due to strong SOC effects.  We have analysed the WAL behavior based on the modified Hikami-Larkin-Nagaoka (HLN) model~\cite{doi:10.1063/1.4773207}. The original HLN model comprises of the phase coherence length, spin-orbit scattering and the elastic scattering terms~\cite{10.1143/PTP.63.707}. In the modified HLN model, Assaf et al.~\cite{doi:10.1063/1.4773207} have shown that the spin-orbit and the elastic scattering terms can simply be approximated by a $B^2$ term and the expression for magnetoconductance $\Delta G (= G(B) - G(0))$ reduces to

\begin{equation} 
\label{Eqn2}
\Delta G (B) = - A \left[ \psi \left(\frac{\hbar}{4 e^2 L_\phi^2 B} + \frac{1}{2} \right) \\ -ln \left(\frac{\hbar}{4eL_\phi^2B} \right)  \right] + \beta B^2,
\end{equation}

%*********************FIGURE 3********************************
\begin{figure}[!]
\includegraphics[width=0.5\textwidth]{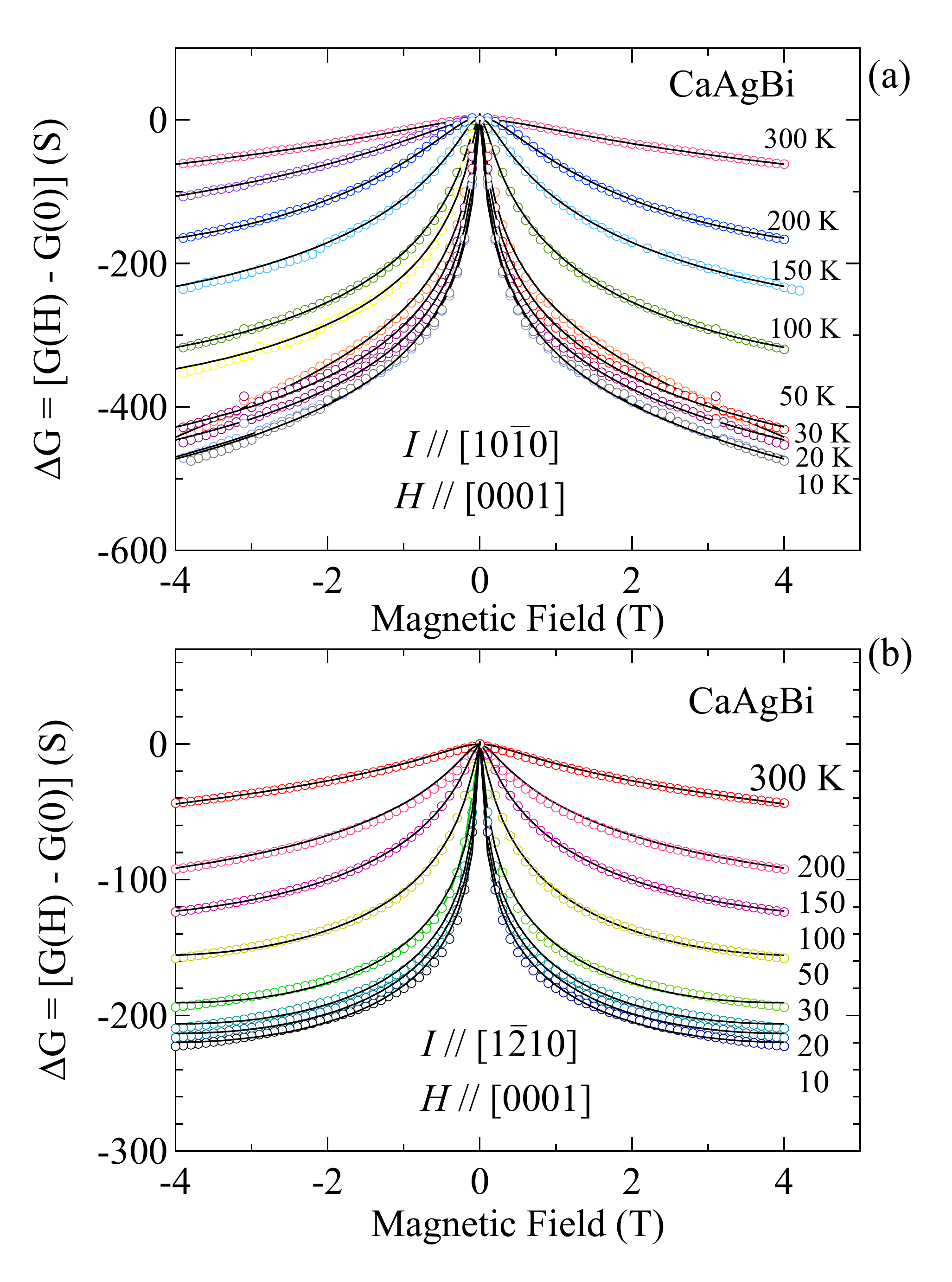}
\caption{(Color online) (a)  Magnetoconductance plot $\Delta G$ $vs$ magnetic field for current parallel to \A100 and \B120 and $B~\parallel$~[0001] direction.   Solid lines are fits to the HLN model given in Eq.~\ref{Eqn2}. }
\label{Fig4}
\end{figure}
%****************************************************************

where the prefactor $A =( \alpha e^2/\pi h)$, $\alpha$ is experimentally found to be $0.5$ for a single coherent conductive channel in a $2D$ electron systems. $\psi$ is the digamma function and $L_{\rm \phi}$ is the phase coherence length and $\beta$ is the coefficient of $B^2$ term which represents additional scatterings. From the measured values of magnetoresistance $\rho_{\rm xx}$ and $\rho_{\rm xy}$, we have estimated the conductivity values using $\sigma_{\rm xx} = \rho_{\rm xx}/(\rho_{\rm xx}^2 + \rho_{\rm xy}^2)$. The plot of magnetoconductance versus field is shown in Figs.~\ref{Fig4}(a) and \ref{Fig4}(b) for current along \A100 and \B120 directions, respectively. The incorporation of the second term leads to a good fit of the HLN model to the $\Delta G$ data even up to high magnetic fields. We have extracted the phase coherence length $L_{\rm \phi}$ and the $\alpha$ parameter from the fitting cure, the results are shown in Fig.~\ref{Fig5}. The value of  $L_{\rm \phi}$ is few hundreds of nm when current is passed along the two principal crystallographic directions in the basal plane. Typically the phase coherence length in disordered metals is of the order of a few thousands of nm~\cite{Lin_2002}. However, in this case, $L_{\rm \phi}$ is relatively smaller due to low mobility as can be inferred from the Hall data, to be discussed later.  It is obvious from Fig.~\ref{Fig5}(a) that the phase coherence length decreases with an increase in temperature. This suggests that the WAL vanishes at a high temperature which is also evinced from the flat magnetoconductance curves at temperatures close to the room temperature. The $\alpha$ parameter remains almost temperature-independent along both the directions thereby suggesting the conducting channels are not altered to a large extent by the variation of temperature. The value of $\alpha$ is much larger than that expected for $2D$ systems. This indicates the possible contribution from the dominated $3D$ bulk channels which are quite often observed in other Bi-based half-Heusler compounds like YPtBi, ScPtBi, HoPdBi etc~\cite{PhysRevB.94.035130, doi:10.1063/1.4936179, Pavlosiuk2016}. However, the obtained value of $\alpha$ is much smaller compared to the Bi-based half-Heusler systems. We tried to analyse the phase coherence length $L_{\rm \phi}$ by considering the electron-electron interaction and the electron-phonon scattering. The expression for $L_{\rm \phi}$ as a function of temperature is given by~\cite{Lin_2002}:

%*********************FIGURE 5********************************
\begin{figure}[!]
\includegraphics[width=0.5\textwidth]{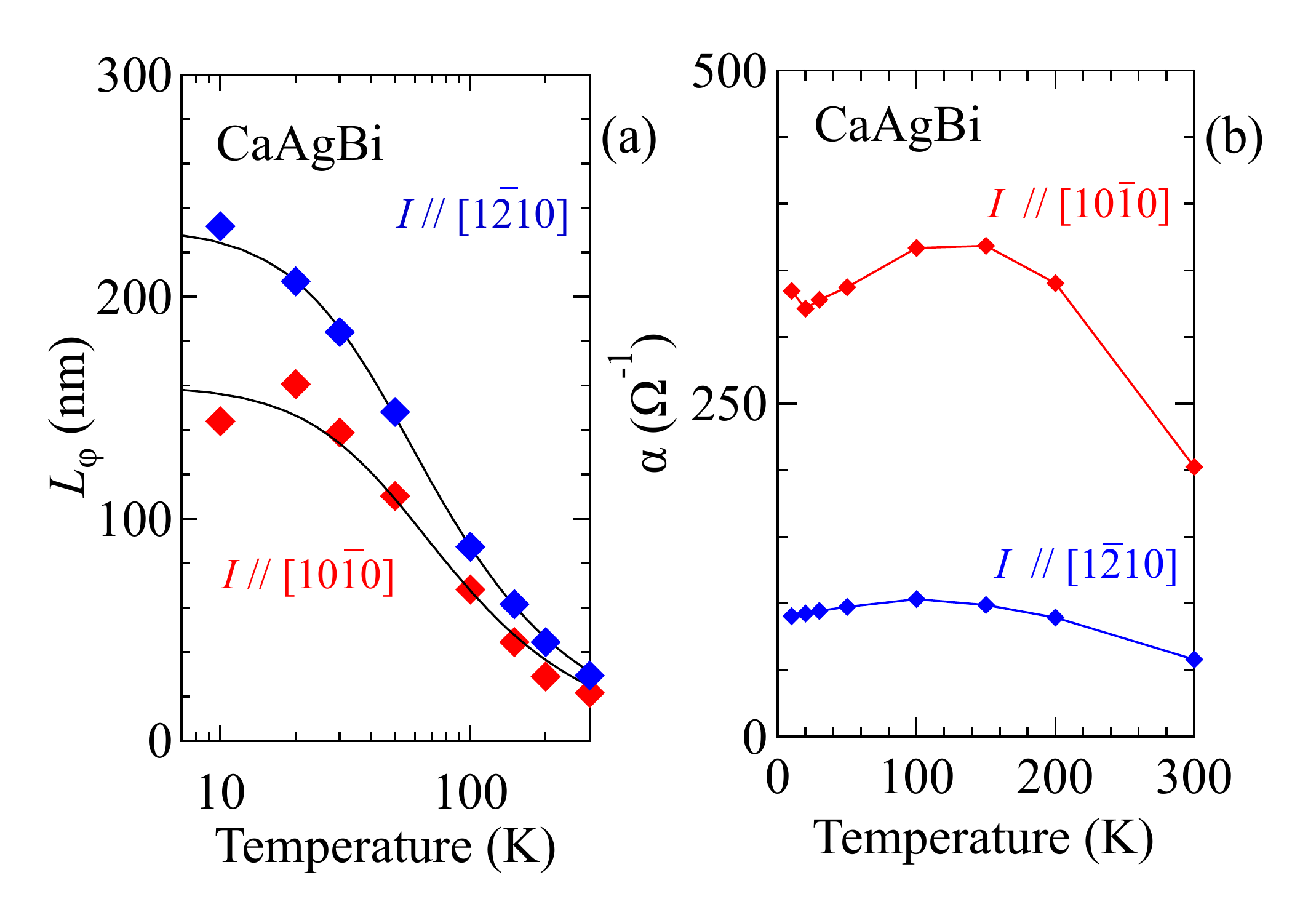}
\caption{(Color online) (a) Temperature dependence of the phase coherence length $L_{\rm \phi}$ as a function of temperature, extracted from the HLN model fitting of the magnetoconductance to Eq.~\ref{Eqn2}. The solid lines are fits to Eq.~\ref{Eqn3}. (b) Temperature variation of the $\alpha$ parameter of Eq.~\ref{Eqn2}. }
\label{Fig5}
\end{figure}
%****************************************************************

\begin{equation}
\label{Eqn3}
\frac{1}{L^2_{\rm \phi}(T)} = \frac{1}{L^2_{\rm \phi}(0)} + A_{ee} T^{el} + A_{ep} T^{ph},
\end{equation}

where $L_{\rm \phi}$(0) represents the phase coherence length at $T = 0$~K, $A_{ee} T^{el}$ and $A_{ep} T^{ph}$ represent the electron-electron and electron-phonon scatterings, respectively. The solid lines in Fig.~\ref{Fig5}(a) show the fitting of Eq.~\ref{Eqn3} to the temperature dependence of $L_{\rm \phi}$ values. The fit has resulted in the values of $A_{ee}$ = 3.4675~$\times$~10$^{-8}$/(nm$^2$K), $A_{ep}$ = 1.761~$\times$~10$^{-8}$/(nm K)$^2$ for $I~\parallel$~\A100 , and $A_{ee}$ = 8.4739~$\times$~10$^{-8}$/(nm$^2$K), $A_{ep}$ = 1.0583~$\times$~10$^{-8}$/(nm K)$^2$ for $I~\parallel$~\B120.  The exponents $el = 1$ and $ph = 2$ with $L_{\rm \phi} = 160$~nm and 231~nm, respectively for $I$ parallel to \A100 and \B120. In the case of three dimensional samples, the predominant scattering mechanism is the electron-phonon scattering while the electron-electron scattering is comparatively weak. However, from our fitting, we find that $A_{ee} T^{el}$ is nearly equal to that $A_{ep} T^{ep}$ (usually, this happens only at very low temperature say less than 1~K)~\cite{Lin_2002}. 

%*********************FIGURE 6********************************
\begin{figure}[!]
\includegraphics[width=0.5\textwidth]{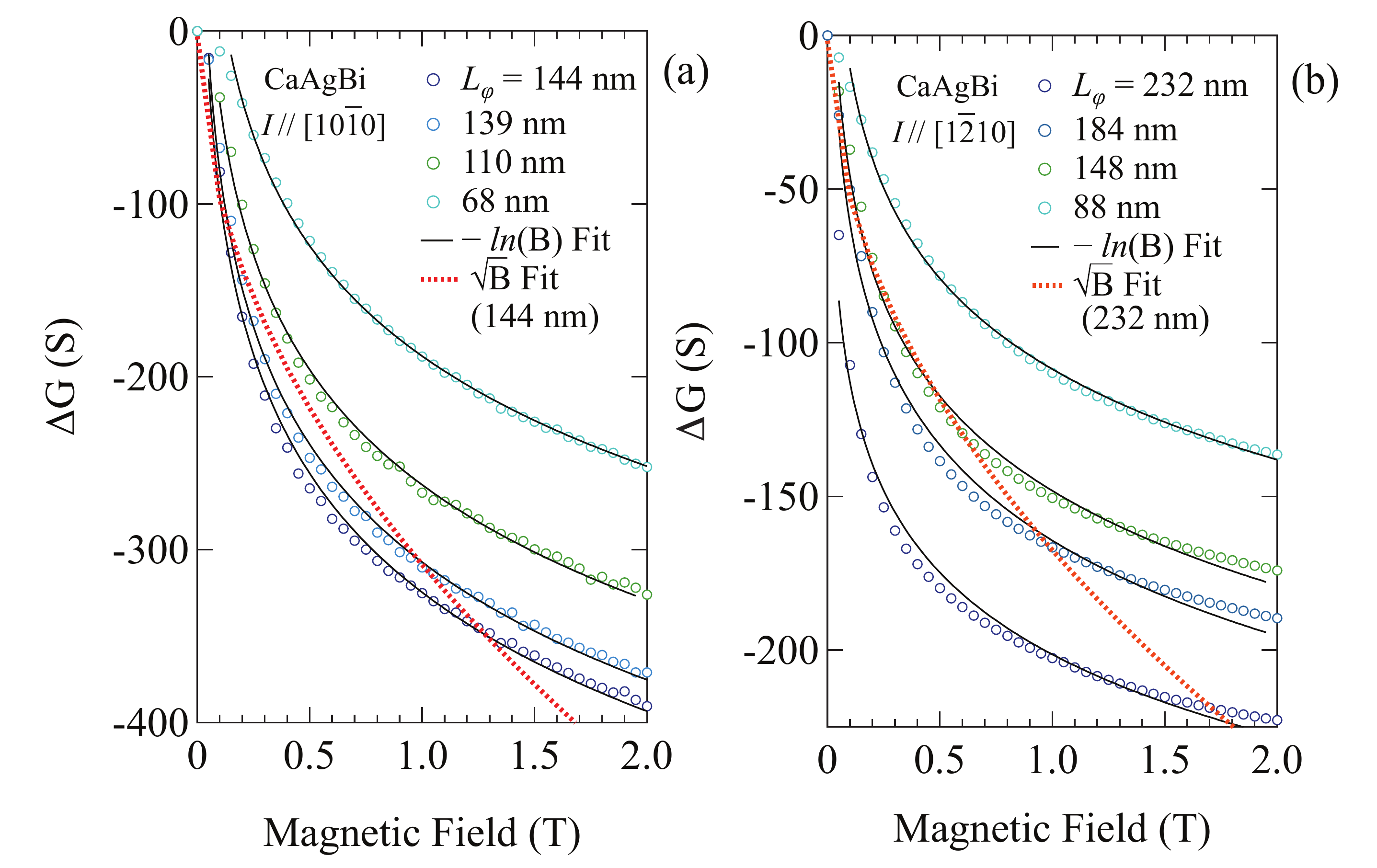}
\caption{(Color online) (a)-(b) Magnetoconductance for different phase coherence lengths when current is passed along the two-principal crystallographic directions on the basal plane. The solid lines are $-$ln(B) fit as predicted by Chen et al~\cite{PhysRevLett.122.196603}. The dashed line shows $\sqrt{B}$ fit to one of the $L_{\phi}$ values for comparison. }
\label{Fig6}
\end{figure}
%****************************************************************

In Fig~\ref{Fig6}, we have plotted the magnetoconductance for different phase coherence lengths ($L_{\phi}$). Recently, Chen et al.~\cite{PhysRevLett.122.196603} have studied the interplay between the effective dimensionality of electron diffusion and band topology in a torus-shaped Fermi surface which encloses a $\pi$ Berry flux inside a nodal loop. Based on the theoretical analysis, they have concluded that weak field magnetoconductance follows $-ln(B)$ scaling behavior for WAL and $\sqrt{B}$ for WL. In Fig.~\ref{Fig6}, we show that the weak field magnetoconductance follows $-ln(B)$ scaling, suggesting the possible weak antilocalization in \CAB. It should be noted that although \CAB~ does not possess a torus-shaped Fermi surface  its weak field magnetoconductance still follows the $-ln(B)$ scaling. For comparison, we have also shown $\sqrt{B}$ dependence of the weak field magnetoconductance for selected values of $L_{\phi}$ which is not matching with our experimental data as shown by dashed lines in Fig~\ref{Fig6}. The scaling law given by Chen et al.~\cite{PhysRevLett.122.196603} obeys the WAL observed in the non-centrosymmetric \CAB~ system. However, it remains to be seen if it can explain other topological systems that do not possess a torus-shaped Fermi surface.

%*********************FIGURE 7********************************
\begin{figure}[!]
\includegraphics[width=0.5\textwidth]{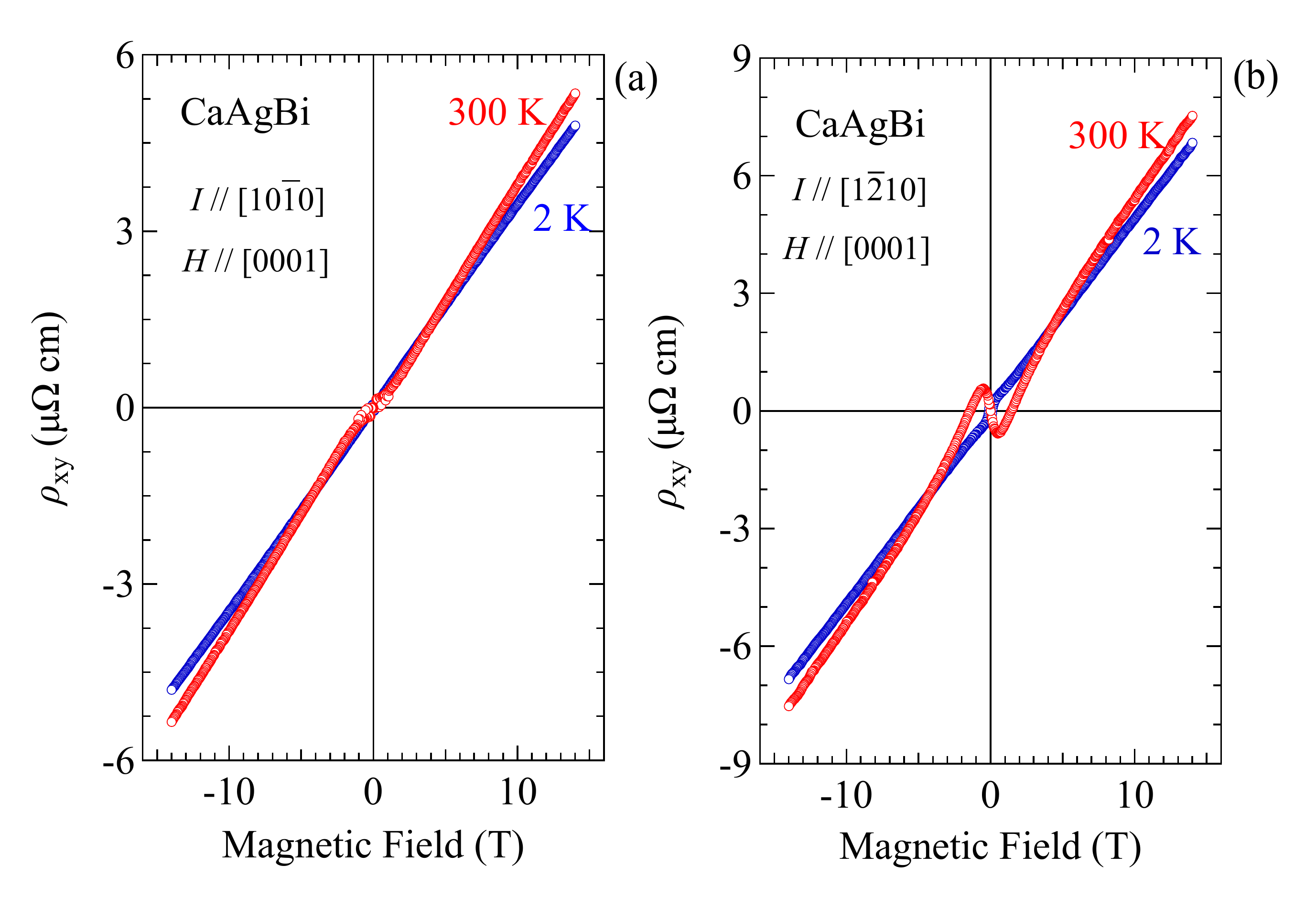}
\caption{(Color online) (a) Hall resistivity $\rho_{\rm xy}$ measured along the two principal crystallographic directions of the basal plane at $T = 2$ and 300~K with the applied magnetic field $H$ parallel to [0001] direction. }
\label{Fig7}
\end{figure}
%****************************************************************

%*********************FIGURE 8********************************
\begin{figure}[!]
\includegraphics[width=0.5\textwidth]{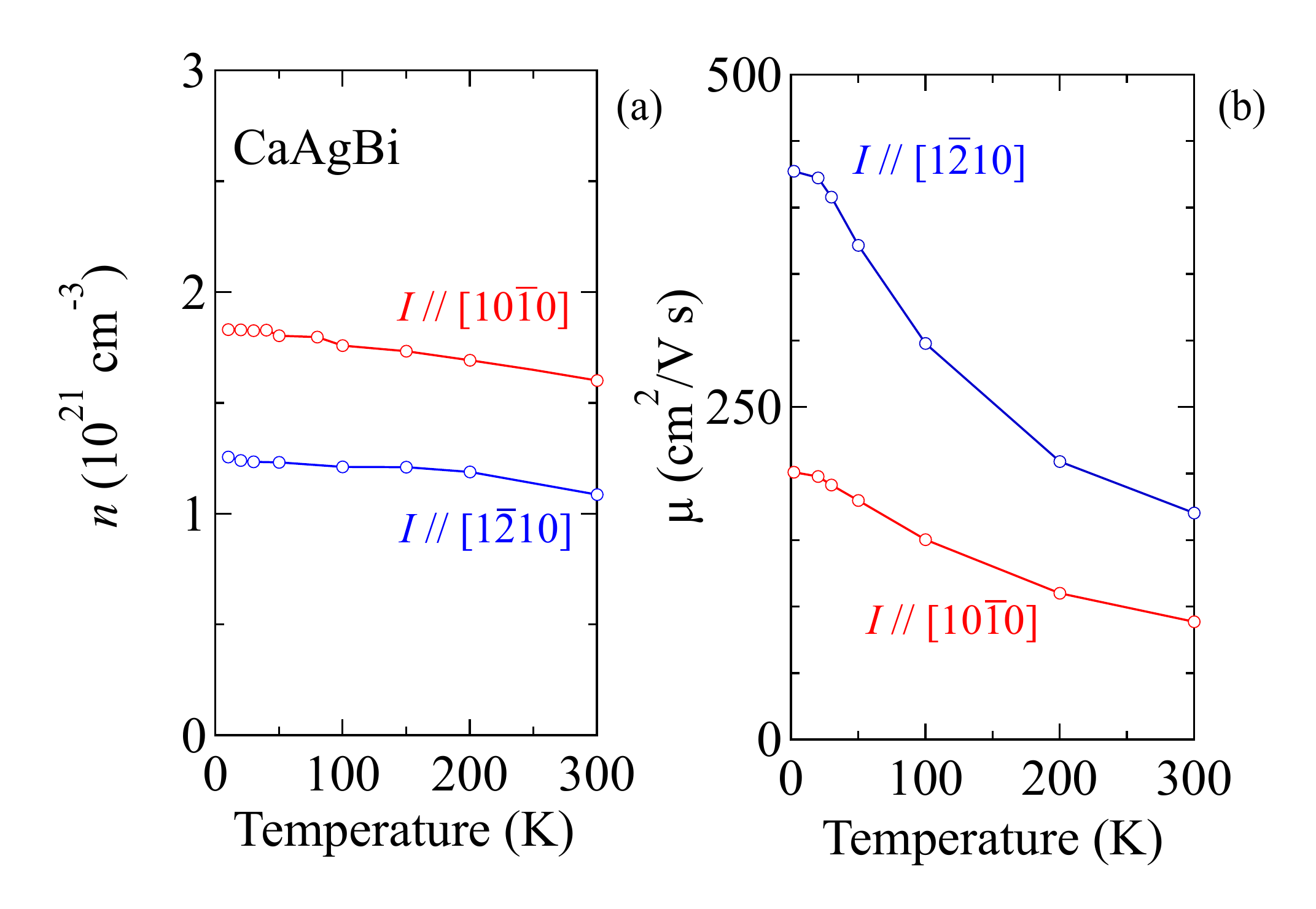}
\caption{(Color online) (a) Carrier concentration as a function of temperature derived from the Hall resistivity data. The subtle difference in the carrier concentration along the two principal basal plane directions is attributed to the error in the sample geometry measurements. (b) Temperature variation of mobility. }
\label{Fig8}
\end{figure}
%****************************************************************

The field dependence of the Hall resistivity $\rho_{xy}$ up to a field of 14~T at two representative temperatures, namely, $T = 2$~ and 300~K, with current parallel to the two principal basal plane directions is shown in Fig.~\ref{Fig7}.  The Hall resistivity shows a negative curvature up to 0.4 T and 1.45~T when the current is passed along the \A100 and \B120 directions, respectively. Beyond this, the Hall resistivity is positive and increases linearly with increasing field,  suggesting that one type of charge carrier is dominant.  The overall behavior of the Hall resistivity remains almost the same with subtle or almost no change for 2~K and 300~K measurements. This kind of behaviour has been experimentally observed in many topological materials, including with Bi$_{1-x}$Sb$x$~\cite{PhysRevLett.111.246603}, ScPtBi~\cite{doi:10.1063/1.4936179}, LuNiBi~\cite{CHEN2019822}. We have estimated the Hall mobility $\mu_{\rm H}$ and the hole carrier density $n$ based on the single band model. The slope of the linear portion of the $\rho_{xy}$ data gives the Hall coefficient $R_{\rm H}$, and the Hall mobility and the carrier concentration can be estimated using the relations: $\mu = R_{\rm H}/\rho_{xx}(0)$ and $n = 1/(e R_{\rm H})$. The plot of the mobility and the carrier concentration as a function of temperature is shown in Fig.~\ref{Fig8}(a) and (b). We find that mobility $\mu_{\rm H}$ ranges from 189 to 83~cm$^2$V$^{-1}$s$^{-1}$ and 44 to 21~cm$^2$V$^{-1}$s$^{-1}$ for current $I$ parallel to \A100 and \B120, respectively. These mobility values are much smaller as compared to the other equiatomic half-Heusler Bi-based compounds. Furthermore, the mobility shows a negative slope with temperature variation, suggesting the increase of lattice scattering as the temperature increases. On the other hand, the carrier concentration is almost temperature-independent over the entire temperature range and lies in the range of 1.8 to 1.0~$\times$~10$^{21}$~cm$^{-3}$, which is higher than the experimentally observed carrier concentration for CaAgAs~\cite{Nayak_2018,doi:10.7566/JPSJ.85.123701}.  This represents the size and shape of the Fermi surface remains unchanged in the entire temperature range for \CAB.  The subtle difference in the carrier concentration values in the basal plane can only be attributed to the error in the sample dimension and/or the voltage probe leads separation.

%*********************FIGURE 8********************************
\begin{figure}[!]
\includegraphics[width=0.5\textwidth]{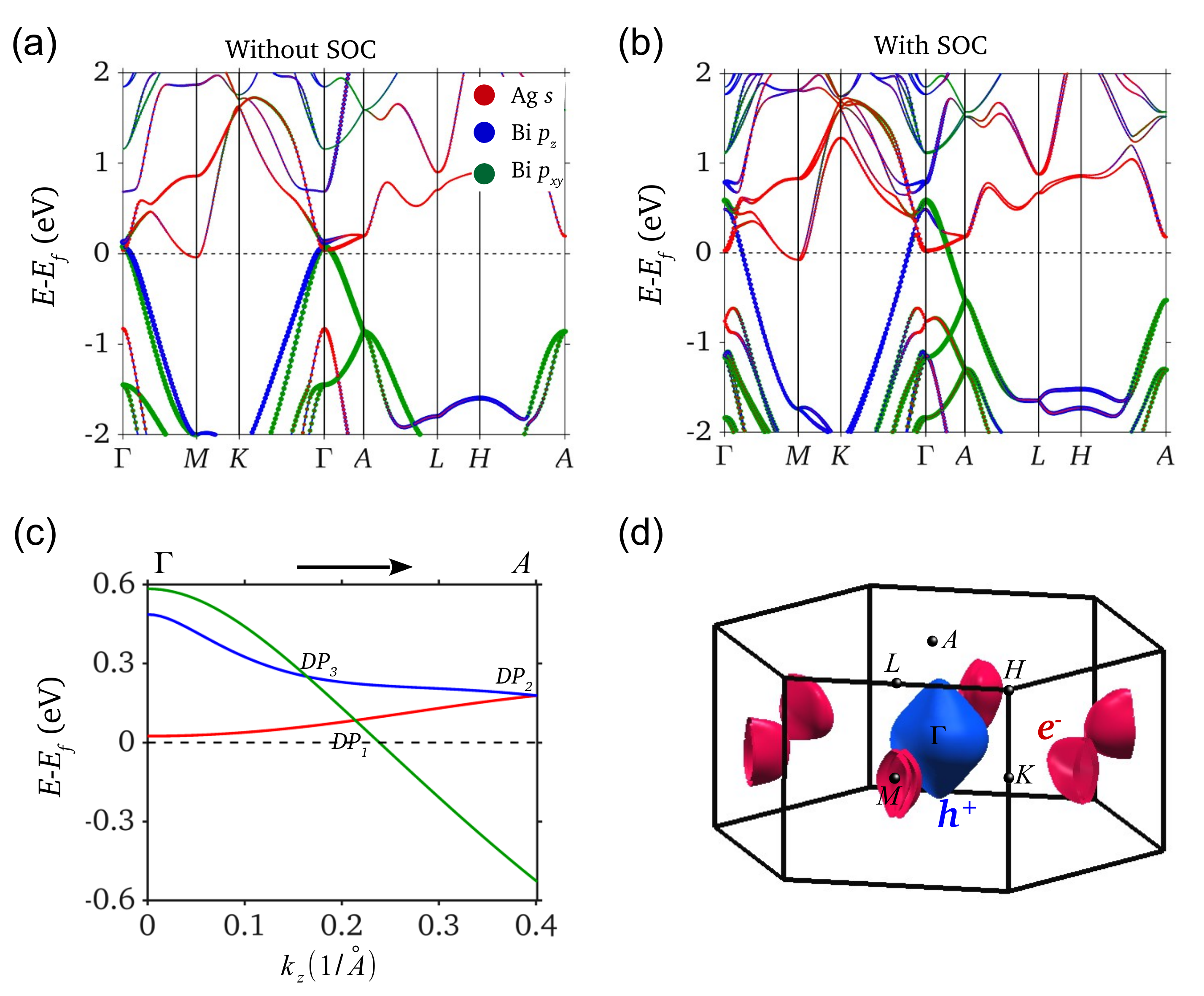}
\caption{(Color online). Electronic structure of \CAB~ (a) without and (b) with spin-orbit coupling (SOC). The contributions of Ag $s$ and Bi $p_{xy}$ and $p_{z}$ orbitals to various bands are proportional to the sizes of various markers. (c) Close-up of the bands along the $\Gamma-A$ direction in (b). $DP_1$, $DP_2$, and $DP_3$ mark three Dirac cones lying on $k_z>0$ axis. (d) Fermi surface of \CAB~ with the electron (red) and hole (blue) pockets. A large hole pocket lies at $\Gamma$ whereas the smaller electron pockets are located near the $M$ point. }
\label{Fig9}
\end{figure}
%****************************************************************

\subsection{Electronic structure analysis of \CAB}

The bulk band structure of \CAB~ along the high-symmetry directions in the bulk Brillouin zone (see Fig. \ref{Fig1}(c)) without and with SOC is shown in Figs. \ref{Fig9}(a) and \ref{Fig9}(b), respectively. It is seen to be a semimetal with both the electron and hole pockets. Without SOC, the valence and conduction bands dip into each other along the high-symmetry directions near the $\Gamma$ point. The orbital resolved band structure shows that low energy bands around Fermi level are derived from Ag $s$ and Bi $p_{xy}$ and $p_z$ states. Usually, Ag $5s$ orbitals have a strong tendency to transfer electrons to $p$ orbitals of Bi and thus lie at higher energy than Bi $p$ states. However, this simplistic picture is inverted near $\Gamma$ point where Ag $s$ derived states lie at lower energy than Bi $p_{xy}$ and $p_z$ states. This suggests a crystal-field induced band inversion semimetal state in \CAB. The inclusion of SOC opens a band gap at the band crossing points along the $\Gamma-M$ and $\Gamma-K$ lines whereas the band crossings stay preserved along $\Gamma-A$ direction [see Fig. \ref{Fig9}(b)]. Regardless, owing to the large SOC, the band inversion strength increases and size of the electron and hole pockets grow larger as compared to these Fermi pockets without SOC as shown in Fig. \ref{Fig9}(d). More specifically, CaAgBi Fermi surface consists of a large biconvex-shaped hole pocket at $\Gamma$ and smaller oval-shaped electron pockets at $M$ point of the Brillouin zone. The larger size of the hole pocket suggests the dominant $p$ type carriers for \CAB. These results are consistent with our measured Hall data which is linear and positive suggesting holes as the majority carriers (Fig. \ref{Fig7}). 

In Fig. \ref{Fig9}(c), we characterize the nodal band crossings along $\Gamma-A$ direction ($k_z$ axis) in the presence of SOC. The three bands derived from Ag $s$, Bi $p_{xy}$, and $p_z$ orbitals cross at three discrete points labelled as $DP_1$, $DP_2$, and $DP_3$. The $DP_1$ and $DP_3$ are generated by accidental band crossings and lie on the rotation axis. On the contrary, $DP_2$ is an essential band crossing point which is mandated by non-symmorphic space group symmetries. Notably, each band on the $k_z$ axis ($\Gamma-A$ line) is doubly degenerate due to the anticommutation relation between $S_{2z}$ and $M_x$ symmetries. Therefore, all three band crossings along $\Gamma-A$ direction realize four-fold degenerate Dirac points. Importantly, the Dirac band crossings are not allowed in the non-centrosymmetric materials due to broken $\mathcal{I}$ symmetry. However, \CAB~ represents a unique noncentrosymmetric system with Dirac points in its electronic structure. It is interesting to mention here that \CAB~ can also support Weyl points but our analysis suggests that they lie at much higher energies than the Dirac points. These results are in agreement with earlier theoretical results on $ABC$-type noncentrosymmetric hexagonal materials\cite{PhysRevLett.121.106404,PhysRevMaterials.1.044201}. 

\section{Summary}

We have successfully grown the single crystal of the non-centrosymmetric \CAB~ and studied its transport properties along the two principal crystallographic directions in the basal plane. Although the magnetoresistance is not high, it exhibits $2D$ WAL effect despite $3D$ nature of the material. It is well documented in the literature that WAL is a quantum interference process which either gets added to or subtracted to from the probability that a charge will return to its initial point. The sign of the quantum correction to the return probability decides WL or WAL. The negative magnetoconductance observed in \CAB~ confirms the WAL in this compound. We show that magnetoconductance can be explained by the modified HLN model. The second term comprising the $B^2$ enables to fit the magnetoconductance data to high magnetic fields. From the analysis of the phase coherence length, we find that the electron-electron scattering is nearly equal to that the electron-phonon scattering which is usually observed at sub-Kelvin temperature range. The weak field magnetoconductance follows the $-ln(B)$ scaling behavior, as predicted by Chen et al.~\cite{PhysRevLett.122.196603} although there is no nodal loop near the Fermi level. Our first-principles calculations show that \CAB~ is a band inversion semimetal with large hole pockets at the Fermi level in agreement with the experimental results. Our results thus suggest that \CAB~can provide an ideal material platform to explore novel physics related to topological semimetals.

\section{Acknowledgement}
We thank Vikram Tripathi, department of theoretical physics, TIFR for useful discussions. This work was supported by the Department of Atomic Energy, Govt. of India.

%\bibliographystyle{Ref.bib}
%\bibliography{Ref_BS}
%\bibliographystyle{prsty_SM}
%\bibliography{Ref_BS_souvik}
%\bibliography{Ref}
\bibliography{CaAgBi_2ndOct2019}

\end{document}